\documentclass[runningheads]{llncs}
\usepackage{graphicx}
\usepackage{balance}
\usepackage{soul}
\usepackage{pifont}
\usepackage{comment}
\usepackage{cite}
\usepackage[title]{appendix}
\usepackage{subfig}
\usepackage[hidelinks]{hyperref}

\AtBeginDocument{%
  \providecommand\BibTeX{{%
    \normalfont B\kern-0.5em{\scshape i\kern-0.25em b}\kern-0.8em\TeX}}}

\usepackage{todonotes}
\newboolean{showcomments}
\setboolean{showcomments}{true}

\ifthenelse{\boolean{showcomments}}
  {\newcommand{\nb}[2]{
  \fbox{\bfseries\sffamily\scriptsize#1}
     {$\blacktriangleright$\textit{\textcolor{red}{#2}}$\blacktriangleleft$}
   }
  }
  {\newcommand{\nb}[2]{}
   
  }
  
\newcommand{\forceindent}{\leavevmode{\parindent=1em\indent}}
\newcommand{\EVExchange}{\emph{EVExchange }} 
\usepackage{graphicx}
\usepackage{listings}
\newcommand{\parag}[1]{\noindent\textbf{#1. }}

\usepackage{orcidlink}
\newcommand{\orcid}[1]{\textsuperscript{\hspace{.1em}\orcidlink{#1}\hspace{.1em}}}

\hypersetup{%
    pdfborder = {0 0 0}
}

\begin{document}
\title{EVExchange: A Relay Attack on Electric Vehicle Charging System}

\author{Mauro Conti\inst{1,2}\orcid{0000-0002-3612-1934} \and 
Denis Donadel\inst{1}\orcid{0000-0002-7050-9369}  \and \\ 
Radha Poovendran\inst{2}\orcid{0000-0003-0269-8097} \and 
Federico Turrin\inst{1}\orcid{0000-0001-5660-2447}}
\authorrunning{M. Conti et al.}

\institute{University of Padova, Department of Mathematics, Italy \and
University of Washington, Seattle, USA \\
\email{\{conti,donadel,turrin\}@math.unipd.it}\\
\email{rp3@uw.edu}}
\maketitle              % typeset the header of the contribution
\begin{abstract}

To support the increasing spread of Electric Vehicles (EVs), Charging Stations (CSs) are being installed worldwide. The new generation of CSs employs the Vehicle-To-Grid (V2G) paradigm by implementing novel standards such as the ISO 15118. This standard enables high-level communication between the vehicle and the charging column, helps manage the charge smartly, and simplifies the payment phase.
This novel charging paradigm, which connects the Smart Grid to external networks (e.g., EVs and CSs), has not been thoroughly examined yet. Therefore, it may lead to dangerous vulnerability surfaces and new research challenges.

\forceindent In this paper, we present \emph{EVExchange}, the first attack to steal energy during a charging session in a V2G communication: i.e., charging the attacker's car while letting the victim pay for it. Furthermore, if reverse charging flow is enabled, the attacker can even sell the energy available on the victim's car! Thus, getting the economic profit of this selling, and leaving the victim with a completely discharged battery.
We developed a virtual and a physical testbed in which we validate the attack and prove its effectiveness in stealing the energy. 
To prevent the attack, we propose a lightweight modification of the ISO 15118 protocol to include a distance bounding algorithm. Finally, we validated the countermeasure on our testbeds. Our results show that the proposed countermeasure can identify all the relay attack attempts while being transparent to the user.

\end{abstract}
\section{Introduction}

The fast growth of Electric Vehicles (EVs) in the market led to the diffusion of new architectures to support the energy demeaning required by the vehicles' battery charging.
Despite the global pandemic, the sales of EVs in the first quarter of 2021 were more than $2.5$ times higher than in the same months of the previous year~\cite{web:EVtrend1}. Furthermore, the International Energy Agency estimates that if governments agreed to encourage the so-called ``Green Transition'', EVs could reach 230 million by 2030.
Vehicle vendors such as Honda plans to convert to electric its entire car production by 2040~\cite{web:honda_electric}. This transition process is also facilitated by the global economic trend, pushing the adoption of renewable energies. The growing concern about the climate crisis leads to a worldwide movement to create a green and sustainable future. In 2018, The United States Environmental Protection Agency estimated that the $28.2\%$ of Greenhouse Gas Emissions in the US is due to the transportation sector~\cite{web:car_pollution}.

With such a forecast on the increase of EVs, the energy request from the electric grid will grow as well. This electric demand increase requires smart management of the charging process of each device to avoid overloads and local blackouts. The most common and upcoming paradigm employed to manage the charging of the EVs is the Vehicle-to-Grid (V2G). V2G systems manage the energy distribution from a Smart Grid to the vehicles (i.e., the final user) by providing a communication channel between the two parties~\cite{Sortomme2011}. It can be used for various features, from the charging schedule during off-peak hours to more advanced services such as automatic authentication and billing. 

V2G is a novel paradigm and, for this reason, it still requires many investigations on security features. When designing such a complex and highly interconnected scenario, security aspects represent extensive and complex requirements, as highlighted by different works~\cite{Antoun2020, Mustafa2013}.
For instance, by exploiting the unique MAC address of a vehicle and unshielded charging cables, it is possible to track a user across different stations~\cite{Baker2019losing}. Since V2G can provide a complete internet connection, the EV is exposed to various threats like malware, affecting the vehicle's internal components. The charging column can be attacked as well, for instance by a denial of service attack, blocking the delivering of the charge service to the users.
Other exploits which have been proved to be effective in the V2G scenario include the profilation of the battery behavior~\cite{sun2020classification} and the profilation of the vehicle charging process~\cite{brighente2021evscout2} based on the electric traces generated from the charging process.

\parag{Contribution} In this paper, we present \emph{EVExchange}, the first relay attack specifically conceived for V2G communication. \EVExchange allows the attacker to exchange the charging flows accounting a victim for the energy consumed. We implemented \EVExchange in both an emulated scenario employing MiniV2G~\cite{miniv2g} and in a physical testbed composed of different Raspberry Pi, proving its functioning and effectiveness. Finally, we propose an extension of the ISO 15118 protocol (i.e., the standard protocol in V2G communication) that utilizes distance bounding to identify relay attack attempts. We tested the distance bounding protocol in both scenarios under different conditions, proving its ability to identify the relay attack.
The contributions of the paper are summarized as follow:

\begin{itemize}
    \item We propose \emph{EVExchange}, the first relay attack conceived for V2G communication.
    \item We implemented \EVExchange in simulated and emulated scenarios based on the ISO 15118 charging protocol standard.
    \item We prove the effectiveness of \EVExchange in stealing the power intended for the victim's car.
    \item We propose a countermeasure which allows to early identify relay attacks such as \emph{EVExchange}. We tested such countermeasure under different scenarios and conditions, proving its effectiveness.
\end{itemize}

\parag{Organization}
The remainder of the paper is organized as follows.
Section~\ref{sec:backgroung} briefly recalls the main concepts useful for the goal of the paper, while Section~\ref{sec:related} provides an overview of the related work. Section~\ref{sec:assimption} outlines the system model and the adversary model assumed. Then, Section~\ref{sec:evexchange} presents \EVExchange attack and its implementation, while Section~\ref{sec:countermeasure} describes the proposed countermeasure.
Finally, Section~\ref{sec:conclusion} concludes the paper with some final remarks.

\section{Background}\label{sec:backgroung}

This section overviews the basic concepts related to the electric vehicle charging system from a communication perspective. In Section~\ref{subsec:v2g_back}, we introduce the V2G paradigm, while in Section~\ref{subsec:iso_back} we analyze the most advanced standard in this field. Then, in Section~\ref{subsec:relay_back} we recall the concept of relay attacks. 

\subsection{Vehicle-To-Grid (V2G)}\label{subsec:v2g_back}
The Vehicle-To-Grid (V2G) concept refers to how an Electric Vehicle can communicate with the power grid. It is a feature reserved for Mode 3 and Mode 4 charges, while Mode 1 and Mode 2 have no communication at all since they employ standard and non-dedicated socket outlets~\cite{VandenBossche2010}. 
The communication can range from simple signaling to high-level communication adopting most of the ISO/OSI layers. 
On the energy side, we can identify two different versions. Unidirectional V2G (also referred to as V1G) employs the communication to manage the charging of the EV smartly. V1G can offer services to the grid, such as load leveling by shifting the power demand to off-peak hours, and the EV owners, by charging the EV when the energy price is lower. This strategy can impact the grid's performances avoiding overloads and local blackouts without requiring huge investments in the infrastructure~\cite{Sortomme2011}.

The bidirectional V2G represents an advanced paradigm. In addition to offering smart management of the charging process, it enables the EV to create a bidirectional power flow with the grid. The discharge of a vehicle can be useful for the grid and the EV's owner in different contexts. The grid can benefit from ancillary services such as frequency regulation and balancing, load leveling, and voltage regulation. On the other side, EV owners can get revenues from the power sold to the grid~\cite{Clement-Nyns2011}.

To support the V2G paradigm, different players proposed different communication protocols. The most widely adopted protocols for the front-end communication between the vehicle and the Electric Vehicle Supply Equipment (EVSE) are ISO 15118, SAE J2847, and CHAdeMO. In the back-end communication between EVSEs and control centers, ISO 61850 and Open Charge Point Protocol are the most used~\cite{Noel2019}. 

In this paper, we uniquely focus on front-end communication. Nowadays, CHAdeMO can be considered the defacto standard. It enables communication through a Control Area Network and does not support any authentication method for the vehicle. However, it is available only on expensive DC chargers, not very suited for private owners. 
SAE J2847 was instead designed for homes. It supports AC and DC charging through Power Line Communication (PLC) communication, and it is suited to manage different technologies, such as smart air-conditioning or smart refrigerators. However, with the expected increase of EVs in the next years, this integration can make it difficult to develop algorithms to manage all the devices smartly.  
The most advanced standard is the ISO 15118~\cite{ISO15118-1, ISO15118-2}. It supports both AC and DC charging and shares the same communication means of SAE J2847, making it possible to employ the same infrastructure partially. Since ISO 15118 can support a vast number of services, ranging from authentication to vehicle's firmware update~\cite{Buschlinger2019}, it aims to be implemented globally and become the standard for the future of electric mobility.

\subsection{ISO 15118}\label{subsec:iso_back}

Firstly released in 2013, ISO 15118 is a modern standard for the regulation of the communications between the Electric Vehicle Communication Controller (EVCC) and the Supply Equipment Communication Controller (SECC). EVCC and SECC are, respectively, the endpoints that manages the transmission on the EV and EVSE~\cite{ISO15118-1, ISO15118-2}. It defines a communication channel via PLC on the Control Pilot (CP) of the IEC 62196 connectors~\cite{iec62196}.

At the beginning of the connection, the Signal Level Attenuation Characterization (SLAC) protocol is employed to pair EVCC and SECC through a series of pulses.
Then, the EVCC broadcasts a default number of UDP packets following the SECC Discovery Request (SDP) protocol to retrieve the IPv6 local-link address of the connected SECC.
After that, the High-Level Communication Protocol (HCP) starts using a TCP communication, generally ciphered using TLS. More information on the packets exchanged can be found in~\cite{miniv2g}.

Unlike the oldest standards (e.g., CHAdeMO), which employ the communication channel only to exchange technical information about the battery and the recharge process, ISO 15118 leverages the high-level communication to provide many services to the grid and the user. The authentication process is based on TLS protocol. The TLS certificate employed for the authentication can be obtained or updated during the connection phase. Payments are managed by the standard which supports External Identification Means such as credit cards, RFID cards, or QR codes. Furthermore, ISO 15118 provides a highly comfortable service called Plug-and-Charge (PnC). This mechanism allows the user to be automatically accounted for the energy requested without using a card or other payment means at the moment of the recharge. In this way, the user only need to insert the plug in his vehicle socket to start the charging process. The PnC authentication mechanism employs the TLS certificate installed in the vehicle and used by the charging system identifies the car~\cite{ElaadNL2018}. The owner can obtain its personal certificate by registering with a charging service provider, as defined in the ISO 15118 standard~\cite{ISO15118-1}.
However, as we will see in this paper, PnC can expose the user to some security threats. 

\subsection{Relay Attacks}\label{subsec:relay_back}

A relay attack is a technique through which an attacker can intercept communication between two entities and replay it in another place in space and time through a proxy~\cite{Hancke2009}. It differs from a Man-in-the-Middle attack since there is no hypothesis that the attacker can understand or modify the information relayed (e.g., communication can be encrypted).

Relay attacks are powerful in many applications, generally in the case of transmission of blocks of independent information or encrypted data. For instance, proximity cards (e.g., credit cards) are a profitable target for relay attacks. In this scenario, the card and the receiver perform mutual authentication, and then all the subsequent traffic is encrypted. Using cryptanalysis to recover the keys might be unfeasible or may require tampering with the hardware with costly instrumentation. An attacker can exploit a relay attack to transfer the entire data flow (including the authentication) from the card to a remote reader. A practical attack consists of relaying the data flow from a victim's credit card to a reader near the attacker to account the victim for the payment. 

\section{Related Works}\label{sec:related}

Although electrical vehicle charging systems are a novel topic, various research papers have examined various aspects of their security. Mustafa \emph{et al.}~\cite{Mustafa2013} proposed a security analysis of the charging system, highlighting different threats for charging at home, at work, or in public places. A similar investigation was conducted by Antoun \emph{et al.}~\cite{Antoun2020} showing possible countermeasures for ISO 15118 and OCPP. Other works addressed specifically the ISO 15118 standard~\cite{Bao2018, Lee2014} proposing threats analysis and security mitigations. However, none of these works analyzed the threats deriving from relay attacks in the charging process or tested the feasibility of the presented attacks in a real or emulated environment.

Few researchers conducted in-depth studies on aspects related to the security of the ISO 15118 standard. Martinovic and Baker showed that it is possible to eavesdrop on the communication between a vehicle and a charging column exploiting the electromagnetic emissions of the PLC on an unshielded cable~\cite{Baker2019losing}.
Hofer \emph{et al.}~\cite{Hofer2013} focused on privacy aspects presenting POPCORN, a protocol that enhances privacy on the ISO 15118 standard. To participate in V2G communication and especially to use PnC, EV should maintain keys and certificates stored inside the vehicle itself. To store these data safely, Fuchs \emph{et al.}~\cite{Fuchs2020hip} designed HIP, a backward-compatible protocol extension for ISO 15118, which enables the generation and storing of keys in a Trusted Platform Module (TPM) within the vehicle. Despite an increasing interest in these security aspects of the standard, to the best of the author's knowledge, there are no available solutions to protect against \EVExchange or similar relay attacks.

There are many scenarios in which relay attacks are used. Its application on Near Field Communication (NFC), for instance, is analyzed in different works in literature~\cite{Lishoy2010nfc, Cavdar2015}.
Recently, researchers have successfully proved the effectiveness of a relay attack on the SARS-CoV-2 contact tracking application, proposing a hashing-based countermeasure to secure the environment without losing privacy~\cite{Casagrande2021tracing}. Also the vehicular environment was interested in this kind of attack:~examples in the literature show possible relay attacks conducted on the passive keyless entry~\cite{francillon2011relay}.
In~\cite{Sani2021} the authors propose a solution to enforce the relay resilience of cryptographic protocols in such application, based on a crypto-chain framework. While there are numerous studies focused on the communication between vehicles and keys, to the best of our knowledge, this is the first study that highlights the threat of relay attacks on a V2G communication.

\section{System and Adversary Models}\label{sec:assimption}

To be successfully implemented, \EVExchange must be performed in a scenario that respects different assumptions from the system and attacker points of view. In this section we outline the system model and we detail the assumption an attacker must respect to implement \emph{EVExchange}.

\begin{figure}%
    \centering
    \subfloat[\centering Without attack.]{
        \includegraphics[width=.45\textwidth]{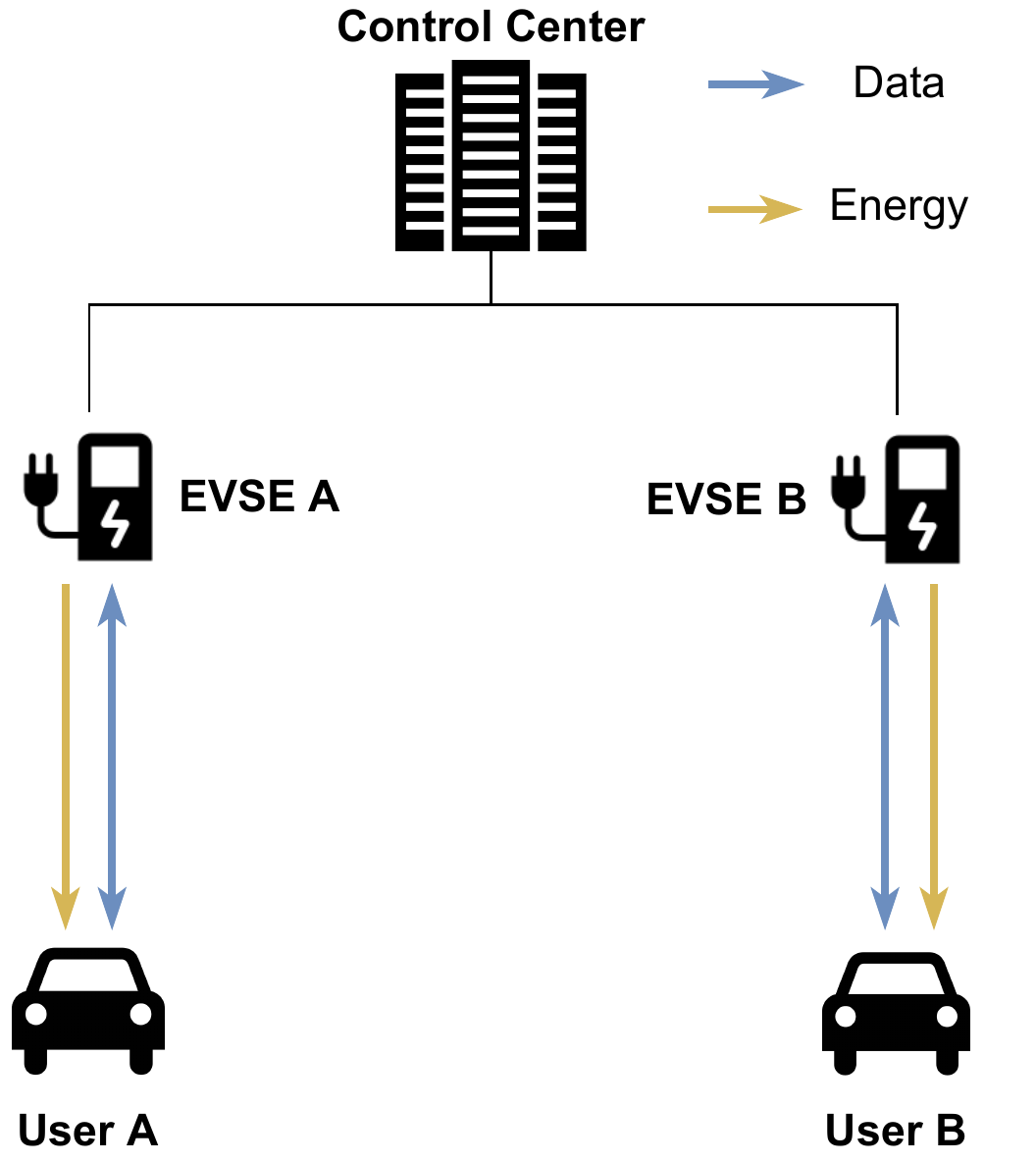}\label{fig:base_scenario}
    }%
    \hfill%
    \subfloat[\centering With \EVExchange attack.]{
        \includegraphics[width=.45\textwidth]{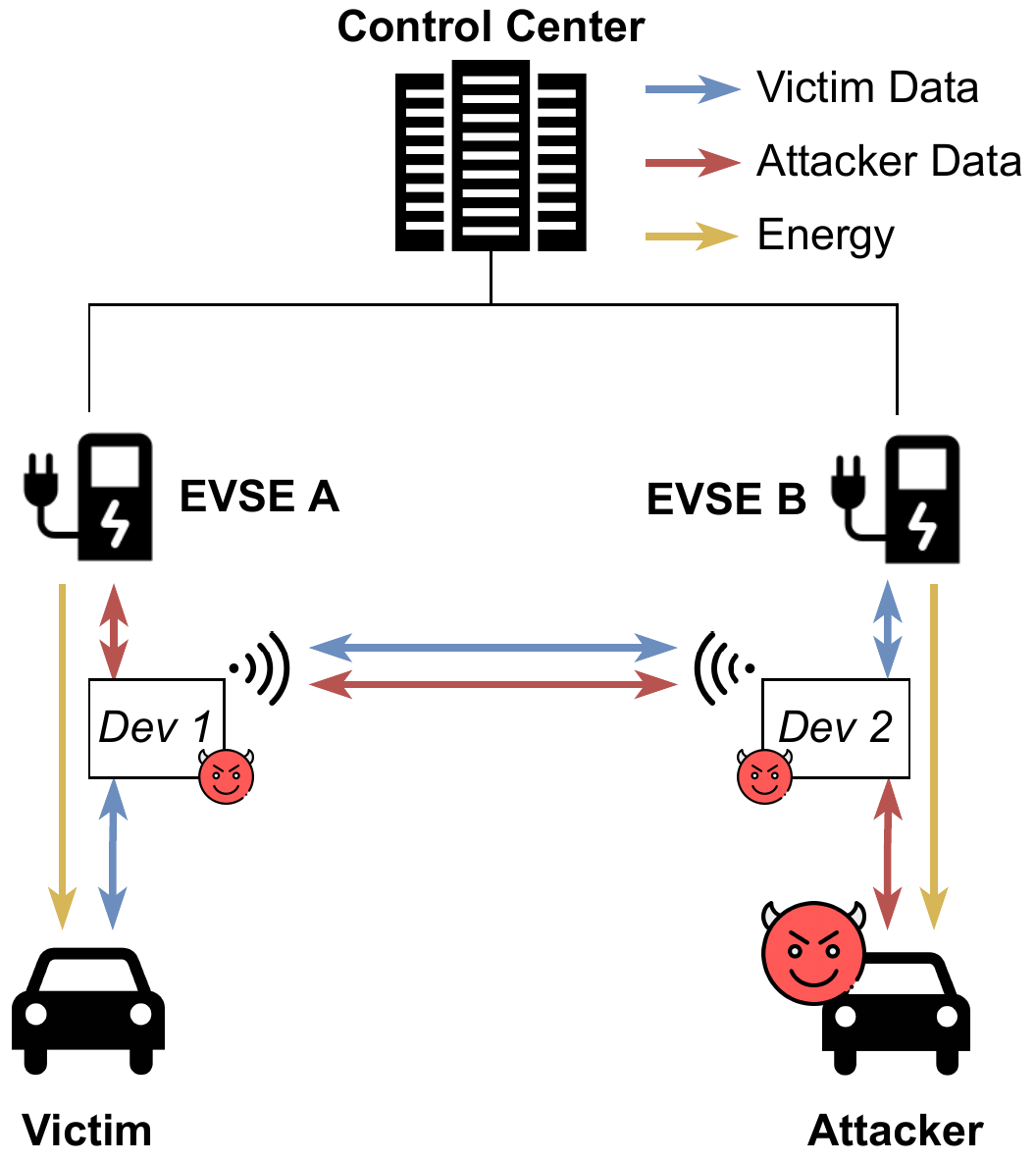}\label{fig:relay_attack}
    }%
    \caption{\centering Scenarios with two EVs charging from two EVSEs connected to the same Control Center (a) and with the malicious devices (b). We represent the Unidirectional V2G scenario for simplicity.}%
    \label{fig:scenarios}%
\vspace{-1em}
\end{figure}

\parag{System Model} Figure~\ref{fig:base_scenario} represents the scenario in which the \EVExchange attack can be performed. As reported in the figure, two EVs are connected to two EVSEs which are in turn managed by the same back-end infrastructures. Since the victim will set the charging parameters used for the attacker's vehicle charge, the attacker must carefully choose two charging columns entirely supported by his vehicle. If more than two EVSEs are available, the attack can be easily extended. However, in this work, we focus on the basic scenario with two EVs and two EVSEs. 
The front-end communication (i.e., between the vehicle and the charging column) employs the most common ISO 15118 standard using the PnC authentication method. Alternatively, this attack is also valid if other means for automatic billing based on a particular ID of the EV are used, such as Autocharge~\cite{web:autocharge} which employs the MAC address of the EV and is commonly used in North Europe. 

EV and EVSE are connected via wired cables, that is the most common setting for power and data, which travel in different cables. Examples of widely employed sockets outlets are Type 1 or Type 2 for AC and Combo 1 or Combo 2 for DC~\cite{DERICIOGLU2018}. There are no substantial differences between them for the purpose of this paper, as soon as the communication is established and billing data are transmitted through the cable in the CP pin.
It can also be possible to extend \EVExchange when wireless communication is employed in the charging process between EV and EVSE. However, we do not consider wireless charging in this work since it is currently rarely used in the real world.

\parag{Adversary Model} 

As a preliminary phase, the attacker must tamper with the charging station to install two malicious devices (i.e., \textit{Dev1} and \textit{Dev2}) as depicted in Figure~\ref{fig:relay_attack}. The two devices can be two simple microcomputers (e.g., Raspberry Pi) with two interfaces to demodulate the PLC in the CP pin and WiFi connection capabilities. A highly skilled attacker could design an ad-hoc device to minimize the device's size to remain undetected. Ideally, each device can be placed in the socket as an adapter, essentially invisible to an average user. Other solutions could be to cut the charging cable to extract the CP cable, cut it and connect it to the two PLC interfaces of the device. The best solution depends on the charging column's type.

Furthermore, the two devices must be connected with each other. While a wired connection is the most reliable and fast solution, it can be visible and could create some suspicion in the user. A wireless connection is the most suited and straightforward approach to avoid this issue. In this work, we employed a standard WiFi connection (i.e., IEEE 802.11ac and IEEE 802.11g) with an intermediate Access Point and in an ad-hoc configuration. 
If the distance between the two devices is significant, high-range wireless connections (e.g., 4G/LTE) can also be employed. 

Once installed and activated, the two devices must block the communication channel between each EV and its legitimate EVSE. Then, they must function as a relay by forwarding the communication coming from an EV to the other device (called \emph{Dev1 to Dev2 relay}, or viceversa), which will recreate the data flow on the EVSE side. 
It is worth noting that the two devices do not need to read the content of the forwarding traffic. This is important because the security standard imposes the usage of TLS to encrypt the communication channel in public places, especially when using PnC~\cite{ISO15118-2}. However, as reported in~\cite{Baker2019losing}, this security measure is often not implemented in practice, exposing the users to many security issues~\cite{miniv2g}. However, even if the traffic is encrypted, the relay process is still feasible, and \EVExchange can be performed.
In this work, we will assume that all the communications between EV and EVSE are always encrypted using TLS. The adversary does not have any valid certificate in addition to the one in the EV. Therefore, it is computationally infeasible for an attacker to decrypt and modify packets on the fly. The attacker is only able to stop and forward the communication flow. 

The key concept to enable \EVExchange attack is that, while the communication flows are forwarded as described above, the energy provided from the two EVSEs is instead directed to the legitimate vehicle (Figure~\ref{fig:relay_attack}). In this way, the attacker can control the energy supplied by the victim's EVSE and vice versa.

\section{EVExchange Attack}\label{sec:evexchange}

After setting the two devices, the attacker can proceed with the \EVExchange attack. We now describe the attack stages through which an attacker can make the victim pay for the energy consumed. We will use Figure~\ref{fig:relay_attack} as reference.

The attacker waits for a victim to arrive at the charging station. When the victim plugs the vehicle into the EVSE A, the attacker will follow by plugging his or her EV into EVSE B. At this point, both users are required to set the charging options they need (e.g., time of departure, energy requirements). Since the two malicious devices are activated, each request made by a user will trigger an action in the EVSE of the other user. 

At this point, to be stealthy, the attacker must replicate the victim's request. However, since the attacker has no clues on the victim's behavior, he can suppose with discrete confidence that the victim will require charging the vehicle since it is the most common operation at charging stations. While it is reasonable to assume that the user will look at the EVSE's display to verify the start of the charging process, the victim probably will not notice a minor difference in the charging parameters, provided that they are displayed in the EVSE. As an example, the forecast duration of the charging process is variable based on the state of charge, the charger type, and the time of the charging. Therefore, it is improbable that an average user can precisely predict this parameter and spot the attack through it.
After requesting the service, since the charging process can take longer, the victim will usually get away from the vehicle to spend the time doing other things while the EV is charging. At this moment, the attacker, who controls the victim's EVSE, can require a stop of charging from the attacker's vehicle. The attacker will now trigger a stop in energy provision in the victim's EVSE (i.e., EVSE A). At the same time, the EVSE connected to the attacker's vehicles (i.e., EVSE B) will continue to follow the victim's request.

Then, when the attacker is satisfied with the charge of the vehicle, he or she can wait for the victims to come back and request a stop of charge for the attacker's EV. Alternatively, the attacker could stop the charging process before the end in his or her charging column to unlock the vehicle and go away, for instance, by using the Emergency Stop button.

Since PnC is employed by the two users in this scenario, the payment of the energy provided to the attacker's EVSE will be billed to to the victim. In the same way, the energy supplied to the victim's EV will be billed for by the attacker but, since the attacker has previously stopped the charge of the victim (at the moment the victim has moved away), he will pay virtually nothing. In contrast, the victim will be billed for a complete charge.

\begin{figure}[h]
    \centering
    \includegraphics[width=.6\textwidth]{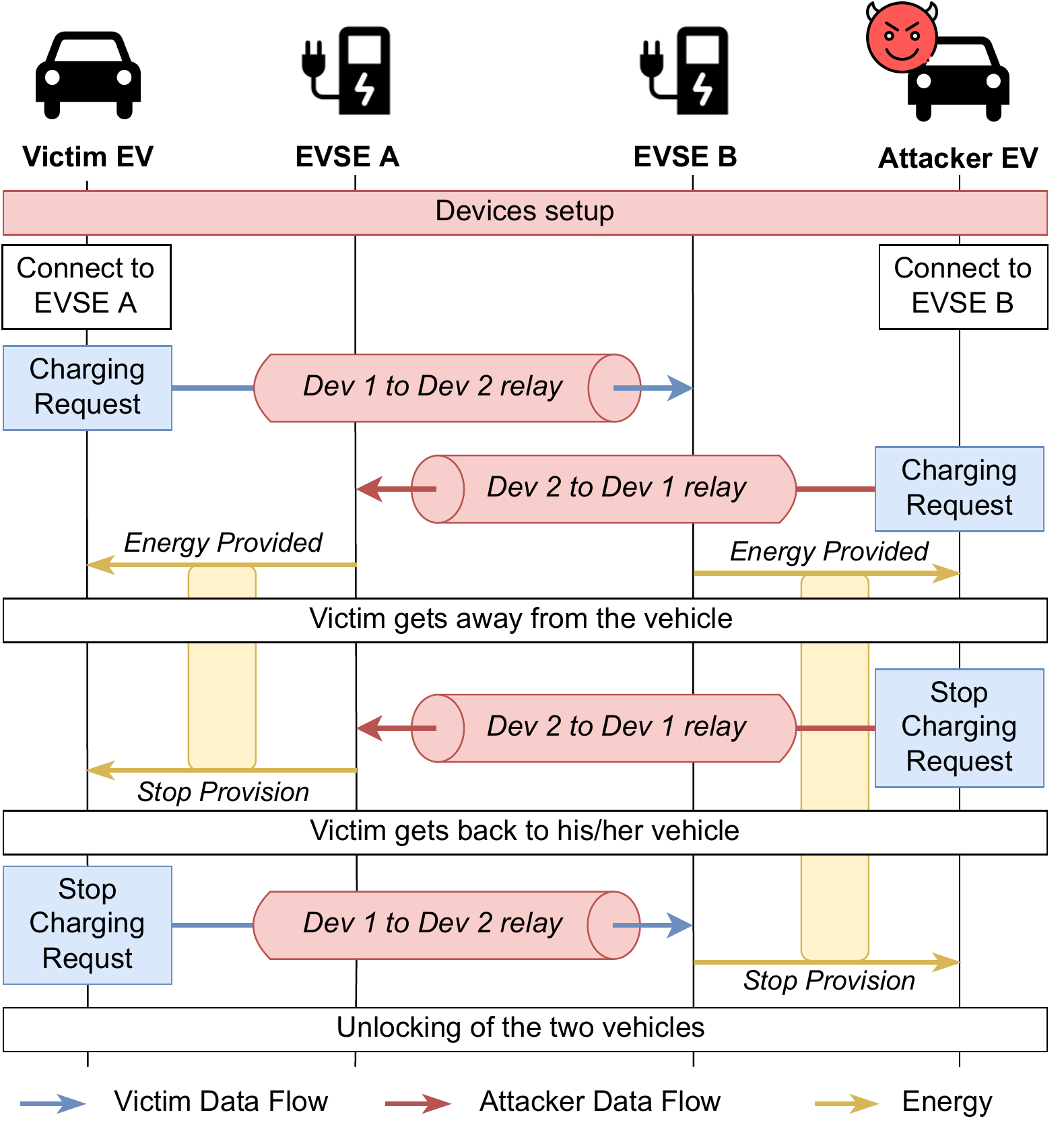}
    \caption{The different phases of the \EVExchange attack. We represent the
Unidirectional V2G scenario for simplicity.}
    \label{fig:attack_phases}
\vspace{-1.5em}
\end{figure}

In the following we summarize the steps of \EVExchange. These steps are also illustrated in Figure~\ref{fig:attack_phases}.

\begin{enumerate}
    \setcounter{enumi}{-1}
    \item The attacker places the two devices as depicted in Figure~\ref{fig:relay_attack};
    \item The victim connects the vehicle to EVSE A; the attacker connects the vehicle to EVSE B;
    \item The two vehicles start a communication with a charging request which is forwarded by the malicious devices; 
    \item The victim, unaware of the attack, goes away from the vehicle; 
    \item The attacker, while recharging by the victim's charging schedule, stops the victim's charge.
    \item When the victim is back, he or she stops the charging process of the attacker.
\end{enumerate}

\vspace{-1em}

\subsection{Variations of the Attack}

\EVExchange attacks can be tailored to achieve different goals. We report here two examples, but many others could be possible.

\parag{Discharge Victim's Battery} We assume a system supporting the bidirectional charge (i.e., the vehicle can sell energy to the grid during peak hours and provide ancillary services to the grid~\cite{Clement-Nyns2011}). In this case, since the attacker controls the victim's communication with the EVSE, he can decide to sell the energy to the grid the power in the battery. Furthermore, by doing so, the revenue will be billed for in the attacker's account.

\parag{Damage Victim's Battery}
One of the most delicate components of the vehicle is undoubtedly the battery. It is subjected to fast degradation through usage, which is responsible for reducing the maximum capacity over time~\cite{Pelletier2017}. In~\cite{brighente2021evscout2} the authors demonstrate the possibility to profile a vehicle based on the battery charging profile.
Some situations can speed up the degradation process, such as extreme operation temperatures, overcharging, and completely draining the battery~\cite{Wu2015charge}. 
Since the attacker controls the victim's charging parameters, he or she can overcharge the battery by requiring energy even if the battery is full. If the bidirectional charge is available, full discharge can be performed as well. Furthermore, an advanced attacker could modify the EVCC or, more simply, modify packets with battery status on the fly to send abnormal charging parameters to the victim's charging column requiring an amount of energy that may damage the battery.

\subsection{Attack Validation}\label{subsec:attack_validation}

The EV charging infrastructure is complex to reproduce and manage since it involves different technical aspects, from the energy to the communication, and includes expensive components. The most common workaround to these limitations is the usage of simulators or emulators. We started our study by testing the attack on implementation of the scenario in MiniV2G~\cite{miniv2g}, an open-source emulator able to simulate networks of EVs and EVSEs. MiniV2G is built on top of Mininet-WiFi~\cite{fontes2015mininet}, a popular software to create realistic virtual networks, running real kernel, switch, and application code. Furthermore, MiniV2G includes RiseV2G~\cite{web:risev2g}, an open-source simulator to implement the ISO 15118 communication. Currently, MiniV2G can only emulate the network communication between EVs and EVSEs without simulating the actual battery charging process. However, this limitation does not affect the implementation of \EVExchange since it is entirely implemented at a network level. For space limitation, we will not discuss the MiniV2G implementation in this work, but we will focus on the development of the physical testbed. However, the MiniV2G implementation and all the code related to this work can be found on Github\footnote[1]{``EVExchange" on Github, github.com/donadelden/evexchange}.

\begin{figure}%
    \centering
    \subfloat[\centering High-level architecture of the testbed.]{
        \includegraphics[width=.47\textwidth]{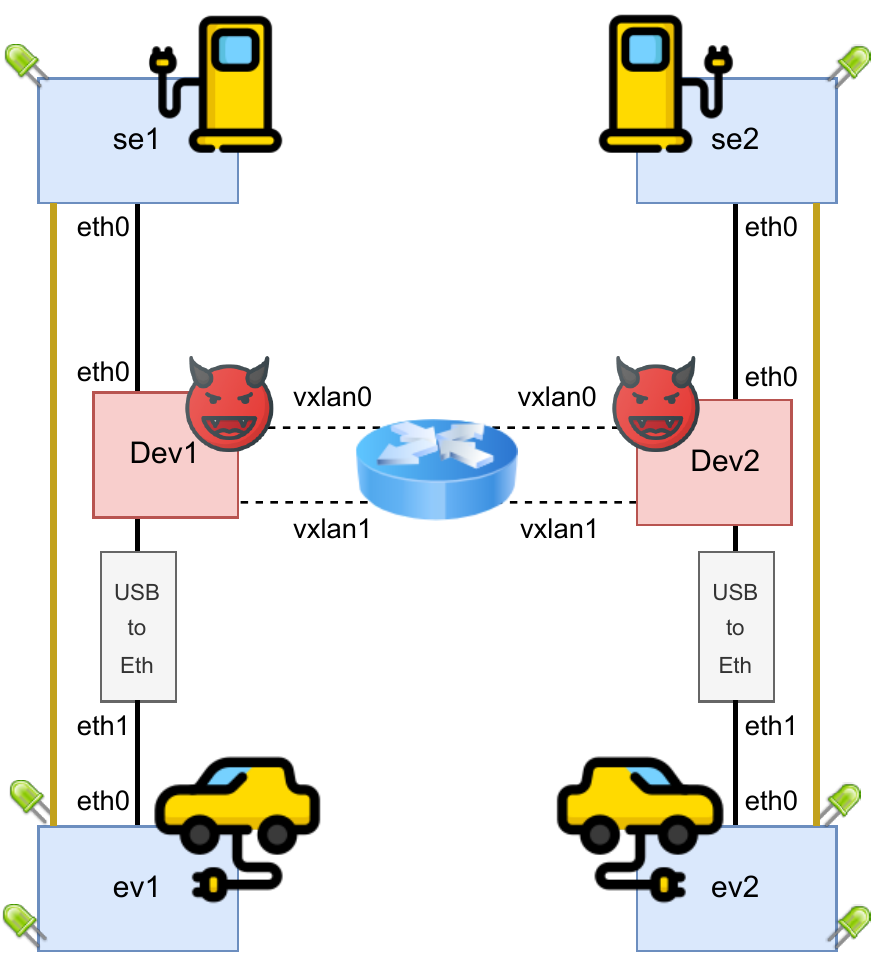}\label{fig:testbed_schema}
    }%
    \hfill%
    \subfloat[\centering A picture of the testbed.]{
        \includegraphics[width=.47\textwidth]{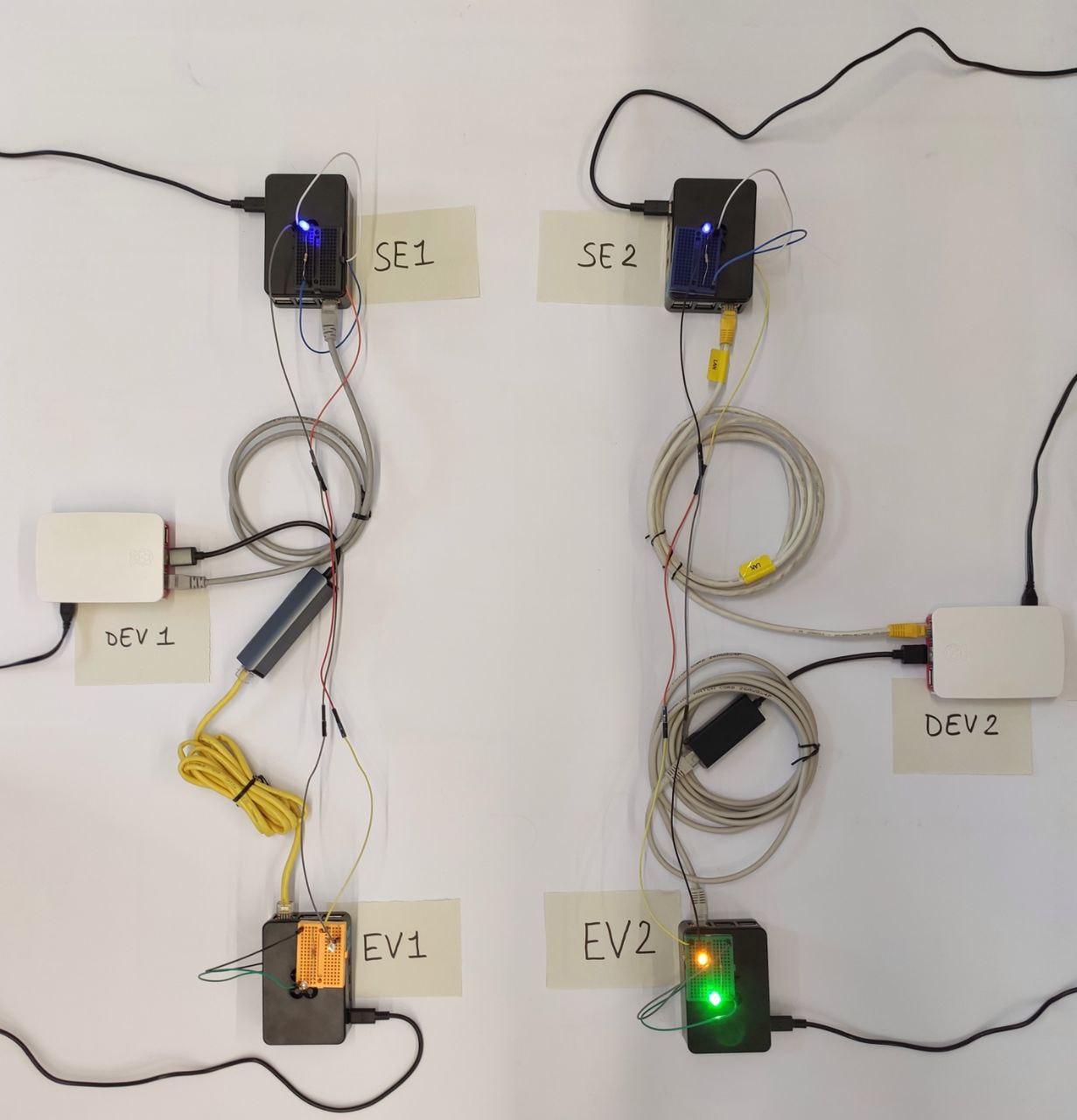}\label{fig:testbed_picture}
    }%
    \caption{\centering The testbed employed to test \EVExchange attack and countermeasure.}%
    \label{fig:testbed}%
\vspace{-1em}
\end{figure}

We preliminary verified the feasibility of \EVExchange on MiniV2G and then we implemented a more realistic scenario by using six Raspberry Pis to emulate vehicles, charging columns, and malicious devices. We used the Ethernet interfaces to simulate the PLC communication while we employed GPIO pins to emulate the energy exchange. We install LEDs to monitor the different stages (i.e., battery charging, energy delivered, authentication completed). As in MiniV2G, we employ RiseV2G in the physical testbed to perform the ISO 15118 communication, with a Python wrapper to turn on the LEDs. Figure~\ref{fig:testbed_schema} represents a high-level schema of the testbed, while Figure~\ref{fig:testbed_picture} illustrates a picture of the testbed developed.

To connect the malicious devices and allow the packets forwarding, we employ Linux \texttt{bridge}~\cite{web:bridge} command to create a channel between the two physical interfaces in each device. These settings do not alter the normal communication flow between EV and EVSE.

When the scripts to activate \EVExchange are executed, bridges are deactivated, and the attack is set up by employing Virtual eXtensible Local Area Network (VXLAN)~\cite{mahalingam2014virtual}. Generally, this tool addresses the need for overlay networks within virtualized data centers accommodating multiple tenants. In our case, we employ VXLANs to create two independent data flows over the wireless network, which can transport packets from one interface of $Dev 1$ to the opposite interface of $Dev 2$. We employ this strategy to configure \EVExchange by relaying data from each EV to the opposite EVSE.

\section{Countermeasure}\label{sec:countermeasure}

To prevent \EVExchange and other potentially related attacks, in this section, we present an extension of the ISO 15118 protocol, which contains a countermeasure based on a distance bounding algorithm. In particular, in Section~\ref{subsec:protocol} we design the distance bounding protocol, while in Section~\ref{subsec:analysis} we discuss the security and the limitation of the proposed algorithm. Then, in Section~\ref{subsec:results}, we describe an implementation of the protocol, providing some numerical results. 

\subsection{Distance Bounding Protocol}\label{subsec:protocol}

To create a countermeasure against \emph{EVExchange}, we can exploit the temporal delay created by the relay process of the communication flows through a wireless channel. The strategy of measuring distance between two devices by considering the Round Trip Time ($RTT$) is known as \emph{distance bounding}~\cite{Brands1994}. As demonstrated in its applications in different contexts in the literature, this approach is the most simple and effective solution to relay attacks. Distance bounding is applied for instance in contactless smart cards~\cite{Drimer2007}, NFC devices~\cite{Henzl2014, Thorpe2020}, and Passive Keyless Entry~\cite{Francillon2012pkes}. This protocol is well suited to work at the application layer in preventing relay attacks since these threats inevitably introduce a measurable delay in the communication.

In general, the distance bounding enables one device (the \emph{verifier}) to securely establish an upper bound on its distance to another device (the \emph{prover})~\cite{Tippenhauer2015}. In our case, the verifier is the victim's $EV$, which wants to check the authenticity of the charging column to which its connected. We consider the EVSE (from now on called supply equipment $SE$ to avoid confusion) as the prover. Therefore, the algorithm's goal is to assess the $EV$ is connected to the correct $SE$ by verifying that the distance between them is no more than an expected value.

The phases of the proposed distance bounding protocol are similar to those proposed by Thorpe \emph{et al.}~\cite{Thorpe2020}, where the authors designed a protocol at the application layer of the NFC protocol. Our algorithm starts after the establishment of the IPv6 connection when the $SE$ starts the listening mode. The core of the proposed solution resides in the fast packet exchange. In this phase, one entity will \emph{immediately} respond to each packet sent by the other. It is possible to compute the $RTT$ precisely and estimate the distance between the two entities from each exchange. In the following, we explain the different phases of the algorithm in detail. Figure~\ref{fig:protocol}, in Appendix~\ref{appendix:count_prot}, graphically summarizes the steps of the protocol. 
\begin{enumerate}
    \item $EV$ generates a random string $\alpha=\{\alpha_1, \alpha_2,\dots,\alpha_k\}$ with a fixed length $k$. Meanwhile, $SE$ generates a random string $\beta=\{\beta_1,\beta_2,\dots,\beta_k\}$ of the same length $k$. These two steps can be done beforehand.  
    \item The fast packet exchange starts for every $i=1,2,\dots,k$ and the $RTT_i$ is measured:
    \begin{itemize}
        \item $EV$ send a UDP packet to $SE$ containing as data the symbol $\alpha_i$;
        \item $SE$ receives $\alpha_i$ and \emph{immediately} responds with an UDP packet including $\beta_i$.
    \end{itemize}
    \item After $k$ exchanges, $EV$ computes the mean $\mu$ and the standard deviation $\sigma$ of the $RTT$s. 
    \item $EV$ compares $\mu$ and $\sigma$ with $\mu_{max}$ and $\sigma_{max}$, which represent the thresholds for $\mu$ and $\sigma$, respectively. If $\mu>\mu_{max}$ or $\sigma>\sigma_{max}$, an error is thrown indicating an attack could be going on. 
    \item If no alert is raised, the secure communication using TLS between the two entities can start as depicted in ISO 15118. Before actually exchanging charging parameters and setting, $SE$ sends to $EV$ the string $S_{SE}=\{\tilde{\alpha_1},\beta_1,\dots\tilde{\alpha_k},\beta_k\}$.
    \item $EV$ computes $S_{EV}=\{\alpha_1,\tilde{\beta_1},\dots,\alpha_k,\tilde{\beta_k}\}$ and compares $S_{EV}$ with $S_{SE}$. If the two strings differ, an alert is raised since an attacker might have forged some packets.
    \item Finally, if no alerts have been raised, the actual charging process can start following the ISO 15118 protocol.
\end{enumerate}

\subsection{Security Considerations}\label{subsec:analysis}

An attacker can employ a series of malicious devices placed in the middle between the EV and the EVSE. For visualization simplicity, in Figure~\ref{fig:protocol}, we represent this set of devices as one single entity called \emph{relay} as a black-box. Considering the adversary devices as a black-box is a reasonable simplification since the legitimate user is unaware of them.
We remark that the \emph{relay} device can selectively or completely relay the traffic flow from two entities as for our hypothesis. Furthermore, the \emph{relay} can eavesdrop on all the not-encrypted communication between the two entities, but it is not equipped with a valid and signed pair of keys to initialize TLS sessions. We do not assume any restriction of the computational capabilities of the adversary. However, it is reasonable to assume that the attacker cannot decipher or modify communication encrypted with TLS.

The proposed distance bounding protocol performs two verifications on the communication. The first one is represented by the effective distance measurement provided by the $RTT$s. The attacker may try to tamper with it by reducing the latency generated by the relay. However, each strategy must be consistent and avoid failure in the second check during the verification of the transmitted data. 

To lower the $RTT$s, an attacker can reduce the relay's complexity by employing, for instance, a faster transmission mode. We exclude the possibility of applying a wired connection since it will be easily spottable by an average user or the service provider. Furthermore, it is common for normal and semi-fast charging stations to be equipped with a detachable cable that must be carried by the driver~\cite{VandenBossche2010}, making even more identifiable a wired relay. An alternative is to employ faster wireless communication modes with respect to the IEEE 802.11 standard, such as 5G, to reduce the protocol overhead and any protocol mode translation. However, this would, on the other hand, increase the system's cost and complexity. For short distances, Bluetooth can be considered, but it will lead to equal or lower performances as WiFi~\cite{Korak2014comparison}. It is worth noticing that the PLC employs HomePlug Green PHY, which has almost no delay at the MAC layer when applied between two entities only~\cite{Chung2006}, making it even harder to create a fast enough channel to avoid detection. Furthermore, it is important to recall that the implementation must be small enough not to draw the victim's attention. 

The previous strategies represent attack optimizations to faster the packet exchange. Another strategy to reduce the $RTT$ could be to tamper with the initial packet flows. Since the initial rapid packet exchange is performed without encryption, the attacker could potentially alter the transmission of the packets. 
For instance, an attacker can decide to send random $\beta_i$ immediately after seeing an $\alpha_i$ to reduce the $RTT$. This process might bypass the first alert control assuring a lower $\mu$ and $\sigma$, but it will be detected during the second control when comparing $S_{EV}$ and $S_{SE}$. By defining $\alpha_i$ and $\beta_i$ values from an alphabet of $N$ symbols, the probability for the attacker to correctly guess the entire string $\beta$ is $\frac{1}{N^k}$. Assuming to employ only the 128 ASCII chars and a sequence of $k=10$ exchanges, we obtain a probability of success for the attacker of $\frac{1}{128^{10}}\approx10^{-22}$ which is negligible. We can further reduce this probability by implementing additional exchanges $k$ and a larger alphabet $N$.

Note that the proposed protocol does not try to prevent \emph{relay} from knowing both $\alpha$ and $\beta$. Instead, it imposes bounds on the \emph{maximum time} by which the information must be received. In other words, when \emph{relay} read the packet containing $\beta_i$, it introduces a delay that makes it too late for the forwarding of the packet to $EV$ and the achievement of a low $RTT$. Furthermore, the transmission of $S_{SE}$ secured by the TLS ensures that \emph{relay} cannot be able to modify it. The only way it is possible to change $S_{SE}$ by an attacker in possession of valid TLS certificates is to pretend to be $EV$ and $SE$ when sending messages to $SE$ and $EV$, respectively. However, we can reasonably assume that the Public Key Infrastructure is solid, and the attacker cannot craft private keys and certificates. 
Nevertheless, it is essential that both the legitimate entities check the validity of their counterpart's certificates before starting the charging process.

\subsection{Evaluation}\label{subsec:results}

To implement the distance bounding algorithms, we wrote two Python scripts to be executed in the EV and the SE, respectively. The protocol starts with a pair of hello messages that enables the EV to get the IPv6 of the SE. Then, the EV starts the algorithm by sending a UDP packet to the SE that acts as a server and immediately responds. This process is iterated 100 times to account for the channel variability.
To evaluate, we compute the mean and the standard deviation of every set of measures. We perform 1000 executions of the described protocol for each scenario to validate the countermeasure.

To verify the feasibility and effectiveness of our countermeasure, we preliminary test it on the MiniV2G emulator under different propagation models and on the physical testbed with different distances between the devices. We report in the following the results related to the physical testbed, and for space limitations, we report in Appendix~\ref{appendix:miniv2g} the result of the MiniV2G emulation.
We create different configurations on the testbed in order to represent different possible scenarios:
\begin{enumerate}
\item A completely legitimate solution, without malicious devices in place (Wired); 
\item A legitimate scenario, with malicious devices inserted but turned off (Wired OFF); 
\item An attack scenario, where the two malicious devices are connected through a cabled Ethernet connection (Wired ON);
\item An attack scenario, where the two malicious devices are connected through a WiFi connection with a router in the middle, placed at 5cm (WiFi 5cm) or 2m (WiFi 2m) from the victim. 
\item An attack scenario, where the two malicious devices are connected through ad-hoc WiFi connection (i.e., without any router in the middle). In this case, we avoid the extra hop between the two malicious devices given by the router (WiFi ad-hoc).
\end{enumerate}

\begin{figure}[h]
    \centering
    \subfloat[\centering Mean $RTT$ ($\mu$)]{
        \includegraphics[trim={40 0 50 0},clip,width=.47\textwidth]{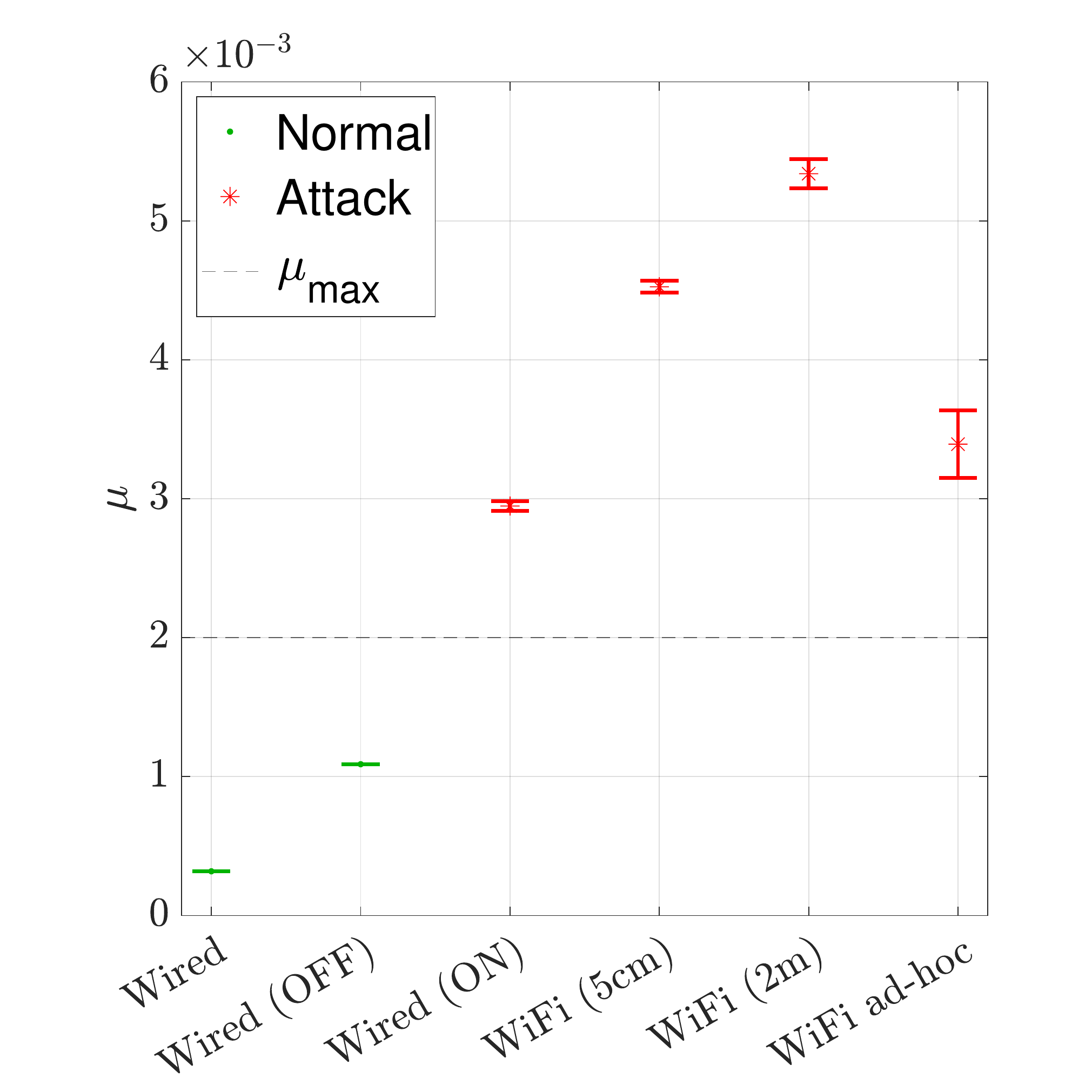}\label{fig:results_mu}
    }%
    \hfill%
    \subfloat[\centering Standard deviation of $RTT$ ($\sigma$)]{
        \includegraphics[trim={40 0 50 0},clip,width=.47\textwidth]{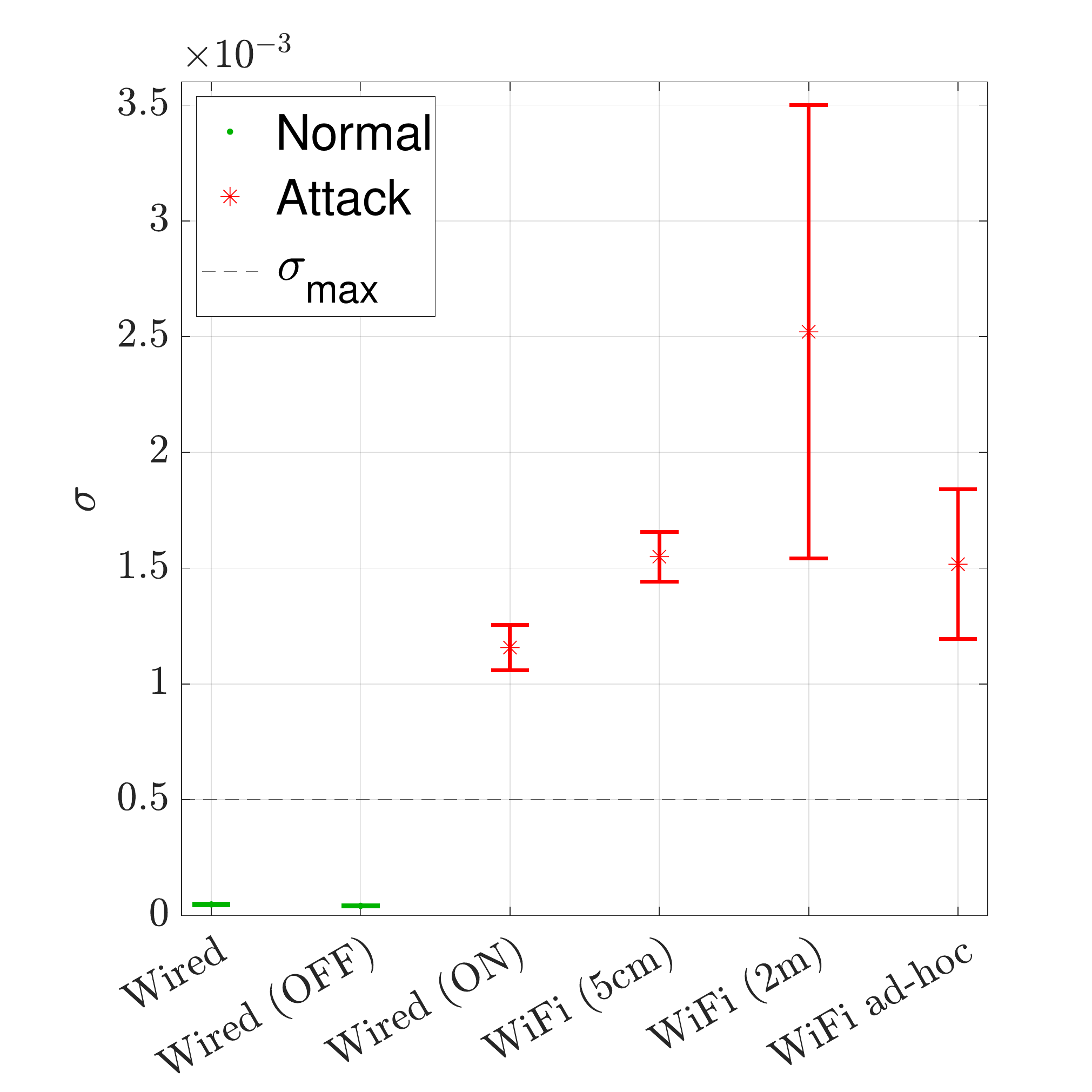} \label{fig:results_std}
    }%
    \caption{\centering $99\%$ confidence interval for the mean (a) and standard deviation (b) in each scenario.}%
    \label{fig:results}%
\vspace{-0.5em}
\end{figure}

We represent the mean $RTT$ in Figure~\ref{fig:results_mu} and the standard deviation of the $RTT$ in Figure~\ref{fig:results_std}. The error bar represents the 99\% percentile. There is a clear separation between the wired data with respect to all the attack cases. This makes it simple to search for good threshold values for $\mu_{max}$ and $\sigma_{max}$, which are represented as a horizontal dashed line. Based on the data we have obtained during our tests, we can safely set $\mu_{max}=2\times10^{-3}$ and $\sigma_{max}=0.5\times10^{-3}$, without almost any risk of having false positives or false negatives.

Note that the time needed for the distance bounding algorithms is generally less than $0.06s$ using 100 fast exchanges, with tops of about $0.3s$ when under attack, which is in practice a rare condition. Furthermore, sufficient security could be ensured even with a few exchanges, reducing the time requirements. Since a charge could last from half an hour to several hours, we can say that extra time added from this countermeasure is negligible and invisible to the end-user. 
We must underline that the experiments were performed in a controlled environment. A thorough evaluation of distance bounding should include a broader spectrum of devices and a wider range of environmental conditions. However, this is beyond the scope of this work.

\section{Conclusions}\label{sec:conclusion}
To support the ongoing diffusion of EVs, the charging process's cybersecurity must be considered to improve users' trust in the system. We demonstrated for the first time that \emph{EVExchange}, a relay attack, is a potent threat against the electric vehicle charging environment against the ISO 15118 protocol.
On one side, \EVExchange can harm the victim, avoiding the charge of its vehicle. On the other side, \EVExchange can damage the EV by exploiting wrong charging parameters and useless charging cycles. Furthermore, \EVExchange allows the attacker to obtain a profit such as free energy and money from the victim. 

To defend against relay attacks, we developed an effective countermeasure able to identify the relay attack in the early stages before sensitive data are shared. The security mechanism adapts distance bounding algorithms to work in the application layer of the ISO 15118 protocol. The countermeasure can always detect the attack in less than $0.3s$ without affecting the normal communication if no attack occurs. 

Since ISO 15118 is a novel protocol, we believe that our work can help the secure development of future versions (such as ISO/DIS 15118-20, under development at the moment of writing~\cite{ISO15118-20}), integrating countermeasures against relay attacks. In future works, the development of novel technology like Wireless Power Transfer could enable a possible extension of \EVExchange to wireless communication between EV and EVSE.

\subsection*{Acknowledgments}
This article has received funding from the European Union's Horizon 2020 research and innovation programme under the Grant Agreement No 825183 for the NGI Explorers project and US Office of Naval Research grant \#N00014-20-1-2636. Denis Donadel is supported by Omitech S.r.l., while Federico Turrin is supported by a grant from the Cariparo Foundation and Yarix S.r.l.. We would like to thank all of them.

\bibliographystyle{splncs04}
\bibliography{biblio}

\begin{subappendices}
\renewcommand{\thesection}{\Alph{section}}%

\section{Distance Bounding Countermeasure}\label{appendix:count_prot}

\begin{figure}[h]
    \centering
    \includegraphics[trim={0 0 3em 0},clip,width=.55\columnwidth]{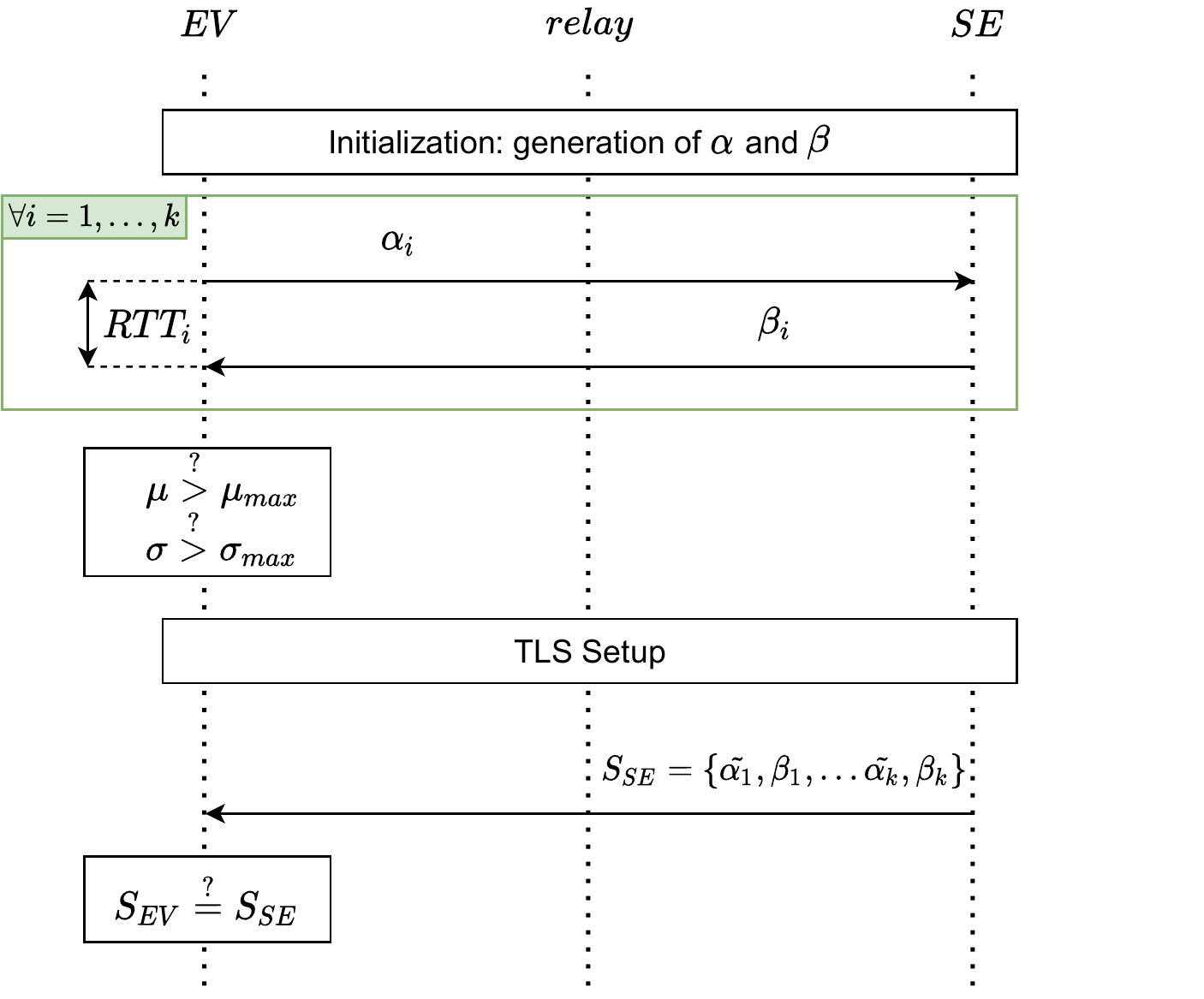}
    \caption{The different steps of the distance bounding protocol.}
    \label{fig:protocol}
\end{figure}

We report in Figure~\ref{fig:protocol} a graphical representation of the Distance Bounding protocol employed as a countermeasure and described in Section~\ref{subsec:protocol}.

\section{MiniV2G Distance Bounding Simulation}\label{appendix:miniv2g}

We report in this section the validation on different scenarios implemented in MiniV2G and performed in a virtual machine with Ubuntu 20.04.2 LTS x64 and 2GB of RAM. The most important parameter that governs the attack's success or failure is the distance between the two malicious devices. We consider two scenarios: two EVSEs at the opposite ends of a parking lot ($10m$) and two adjacent parking spots ($2m$). To emulate a wireless connection in the emulator, we employ different propagation models included in Mininet-WiFi~\cite{fontes2015mininet}. 

We chose as possible models Log Distance Path Loss (LDPL) and Log Normal Shadowing (LNS), both with $exp=2$. As presented in~\cite{Akhtar2019}, these two models are suited to simulate a connection in free space and urban area. Furthermore, we test with two different WiFi versions (i.e., IEEE 802.11g and IEEE 802.11ac).

\begin{figure}%
    \centering
    \subfloat[\centering Mean $RTT$ ($\mu$)]{
        \includegraphics[trim={25 0 50 0},clip,width=.45\textwidth]{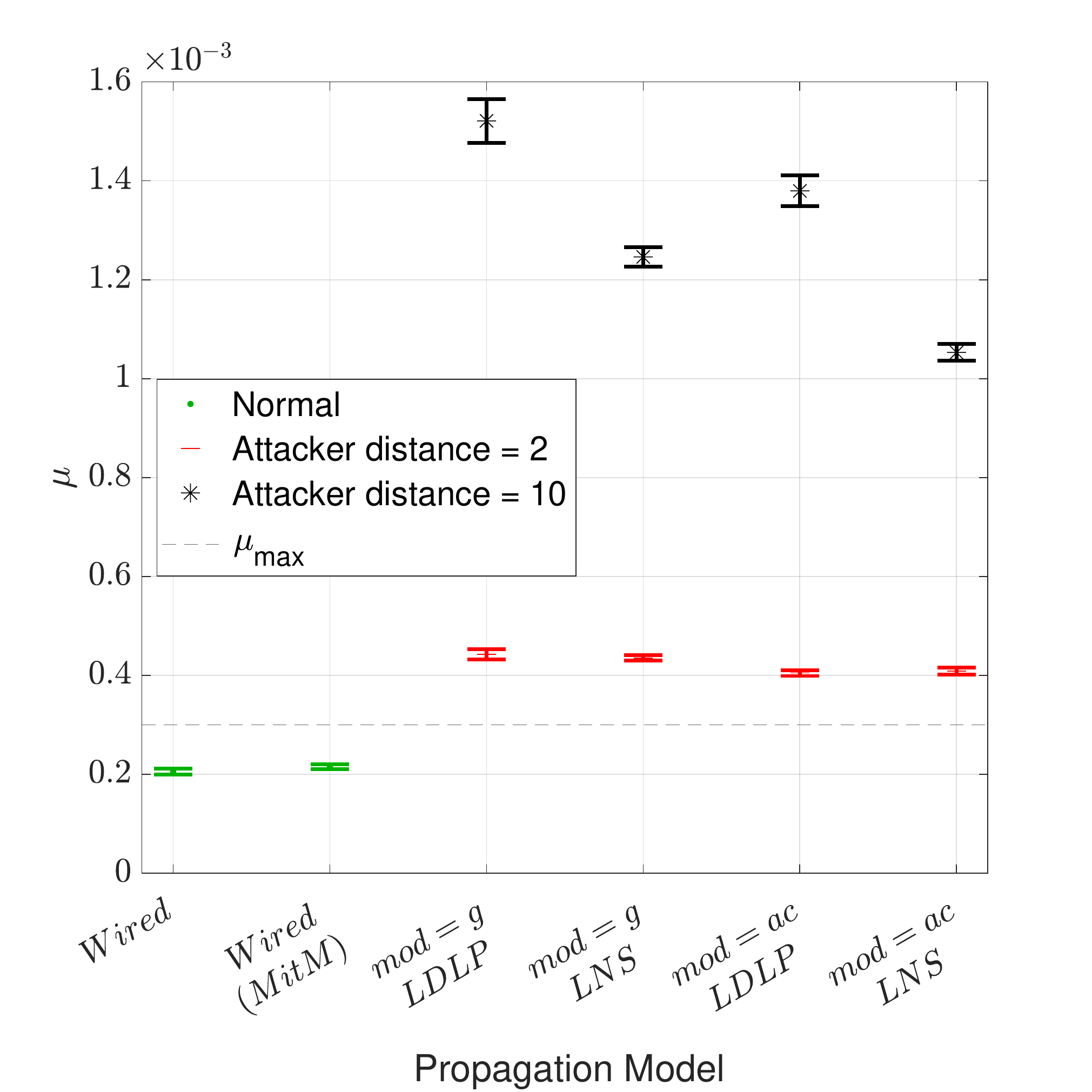}\label{fig:results_mu_mini}
    }%
    \hfill%
    \subfloat[\centering Standard deviation of $RTT$ ($\sigma$)]{
        \includegraphics[trim={25 0 50 0},clip,width=.45\textwidth]{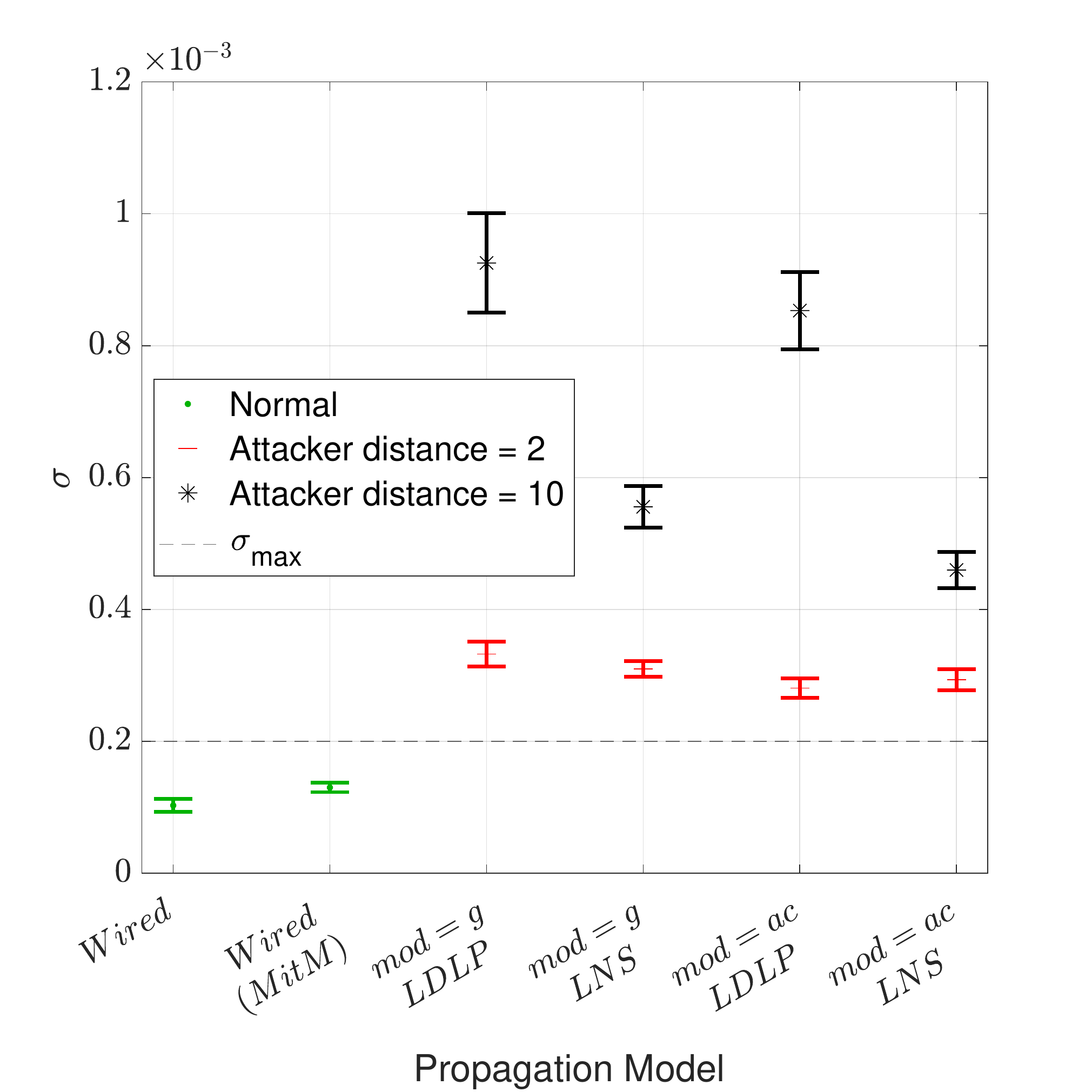} \label{fig:results_std_mini}
    }%
    \caption{\centering  99\% confidence interval for the mean (a) and standard deviation (b) in each different scenario emulated using MiniV2G.}%
    \label{fig:results_mini}%
\vspace{-1em}
\end{figure}

We represent the mean $RTT$ in Figure~\ref{fig:results_mu_mini} and the standard deviation of the $RTT$ in Figure~\ref{fig:results_std_mini}. As in the data presented in Section~\ref{subsec:results}, the error bar represents the 99\% percentile, and there is a clear separation between the wired data and all the other malicious cases. 

\end{subappendices}

\end{document}